\documentclass{article}

\usepackage{arxiv}

\usepackage[utf8]{inputenc} 
\usepackage[T1]{fontenc}    
\usepackage{hyperref}       
\usepackage{url}            
\usepackage{booktabs}       
\usepackage{amsfonts}       
\usepackage{nicefrac}       
\usepackage{microtype}      
\usepackage{lipsum}
\usepackage{graphicx}
\usepackage{epstopdf}
\title{Spectroscopy on nanoparticles without light}

\author{J.~Fiedler~\thanks{Centre for Materials Science and Nanotechnology, Department of Physics, University of Oslo, P. O. Box 1048 Blindern, NO-0316 Oslo, Norway}\\
Physikalisches Institut\\Albert-Ludwigs-Universit{\"a}t Freiburg\\Hermann-Herder-Str. 3\\79104 Freiburg, Germany\\
\texttt{johannes.fiedler@physik.uni-freiburg.de}
\And
C.~Persson\\Centre for Materials Science and Nanotechnology\\Department of Physics\\University of Oslo\\P. O. Box 1048 Blindern\\NO-0316 Oslo, Norway
\And
S.~Y.~Buhmann~\thanks{Freiburg Institute for Advanced Studies, Albert-Ludwigs-Universit{\"a}t Freiburg, Albertstr. 19, 79104 Freiburg, Germany}\\
Physikalisches Institut\\Albert-Ludwigs-Universit{\"a}t Freiburg\\Hermann-Herder-Str. 3\\79104 Freiburg, Germany
}

\begin{document}
\maketitle

\begin{abstract}
One of the most important tools in modern science  is the analysis of electromagnetic properties via spectroscopy. The various types of spectroscopy can be classified by the underlying type of interactions between energy and material. In this paper we demonstrate a new class of spectroscopy based on Casimir interactions between a solid investigated object and a reference surface embedded in an environmental liquid medium. Our main example is based on the measurement of Hamaker constants upon changing the concentration of an intervening two-component liquid, where we demonstrate a possible reconstruction algorithm to estimate the frequency-dependent dielectric function of the investigated particle.
\end{abstract}


\section{Introduction}
Spectroscopy in general is a method to investigate a material's properties, in particular its electromagnetic properties for the characterisation of an unknown material with respect to its constituents~\cite{AAS98} (type and stoichiometry) and its optical behaviour~\cite{Graner_d_2019} (such as band gaps in semiconductors). A huge challenge is the experimental determination of the dielectric function $\varepsilon(\omega)$ of a material which is due, on the one hand, to the large  range of the electromagnetic spectrum and, on the other hand, to the function being complex-valued. In principle, the measurement of the dielectric function ensues by the excitation of the system at the investigated frequency and measurement of its response. The complexity of such experiments is due to the fact that each spectral range requires a different source for the exciting beam. For instance, measurements in the low frequency range (radio waves), where the electromagnetic wave couples to the states of the core spin, require nuclear magnetic resonance spectroscopy~\cite{PhysRev.57.522,PhysRev.57.111,PhysRev.69.37}, which is followed by micro waves, coupling to the electronic spin states and rotational states, which are investigated via electron paramagnetic resonance spectroscopy~\cite{zavoisky1946paramagnetic}. The next spectral range is investigated by Terahertz spectroscopy~\cite{Junginger:10,BUCHNER199957}. Before reaching the visible range, the infrared spectrum has to be investigated via infrared~\cite{doi:10.1002/andp.18852600305} and Raman spectroscopy~\cite{Wolf_1996}, for instance. In the visible and ultraviolet ranges, ultraviolet-visible spectroscopy~\cite{doi:10.1002/biot.201200153} or reflection spectroscopy~\cite{KOENIG199977} is typically used. Finally, X-ray spectroscopy and gamma spectroscopy~\cite{Gordon08} complete the spectrum. To conclude, a huge number of different experiments has to be performed to measure the complete dielectric function of a material. The dielectric function being complex poses an additional challenge. For instance, via reflection spectroscopy one usually obtains the real part of the refractive index $n$ and/or the attenuation coefficient $k$, which is connected with the absorption coefficient. The combination of both results is required to infer the dielectric function. However, often both ingredients are often not simultaneously available and one needs to apply the Kramers--Kronig relation~\cite{deL.Kronig:26,kramersdiffusion}, which requires the knowledge of the complete spectrum for the imaginary or real part to calculate the respective other.

\begin{figure}[t]
    \centering
    \includegraphics[width=0.5\columnwidth]{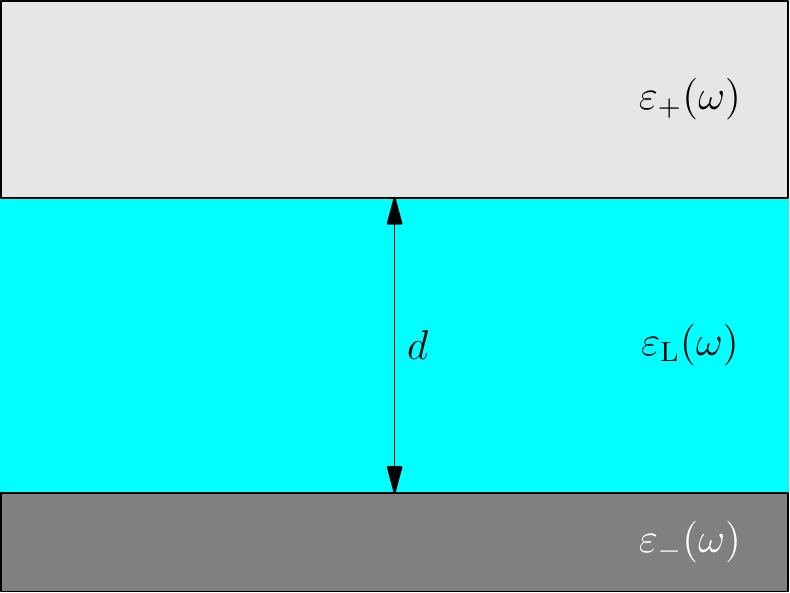}
    \caption{Sketch of Casimir force experiments via atomic force microscopy. A sphere with dielectric function $\varepsilon_+$ and a large curvature radius is attached to a cantilever, which is closely separated from a substrate $\varepsilon_-$. By varying the distance between both objects the Casimir force interacting between them can be measured. In addition to the vacuum experiments a liquid $\varepsilon_{\rm L}$ surrounds the particle.}
    \label{fig:C-AFM}
\end{figure}

However, as we wish to elaborate here, alternatives exist in the form of experiments with a good accuracy whose results directly depend on dynamical dielectric functions, for instance, Casimir experiments via atomic force microscopy in vacuum~\cite{PhysRevA.82.062111,doi:10.1021/la4008558} and in medium~\cite{PhysRevA.81.062502}. Figure~\ref{fig:C-AFM} illustrates the typical experimental setup. A spherical nanoparticle is attached to the cantilever of the atomic force microscope. The radius of the sphere should be large enough to neglect effects caused by the curvature of the surface. By changing the distance $d$ between the sphere and the substrate the Casimir force can be measured. In such experiments, one typically uses a sphere of large curvature radius to overcome the issue of arranging two parallel plates and applies the Lifshitz formula for the theoretical description.

In this manuscript, we consider the above mentioned Casimir experiment and show a possible way to extract the dielectric function of one plate from the measured Casimir force data. In order to vary the Casimir force, we assume a liquid surrounding the sphere which consists of two components. Hence, the observed force depends on the concentration of the two liquids. We consider a simple example to illustrate the method with a textbook inversion procedure to infer the dielectric constant from the force data. The method can easily be extended beyond this proof of principle demonstration to improve the precision or generalise to other measurement setups.

\section{Theory and reconstruction method}

\subsection{Casimir effect in media}
The Casimir effect, the typically attractive force between two dielectric plates with permittivities $\varepsilon_-$ (bottom plate) and $\varepsilon_+$ (top plate) separated by a distance $d$ with a medium permittivity $\varepsilon_{\rm L}$, can be described via the sum over the exchange of virtual photons between them~\cite{PhysRevA.79.027801,Buhmann12a,Friedrich}
\begin{eqnarray}
 f(d) = -\frac{k_{\rm B}T}{\pi}\sum_{m=0}^\infty {}' \int\limits_0^\infty \mathrm d k^\parallel \, k^\parallel \kappa^\perp \sum_{\sigma=s,p}  \frac{r_\sigma^+r_\sigma^-\mathrm e^{-2\kappa^\perp d}}{D_\sigma}\, ,
\end{eqnarray}
with the Boltzmann constant $k_{\rm B}$, the temperature $T$, the primed sum denotes that the first term has to be weighted by 1/2, the Matsubara frequencies $\xi_m = m \, 2\pi k_{\rm B}T/\hbar$, 
the multiple reflection terms $D_\sigma = 1-r_\sigma^+r_\sigma^-\mathrm e^{-2\kappa^\perp d}$ and  the imaginary part of the transverse wave vector $\kappa^\perp_i = \sqrt{{k^\parallel}^2 + \varepsilon_i \xi^2/c^2}$. The coefficients $r_\sigma^\pm$ are the generalised reflection coefficients at the interfaces for $p$- and $s$-polarised waves~\cite{Tomas1995,Chew}
\begin{equation}
 r_{p}^{i,j} = \frac{\varepsilon_j \kappa^\perp_i - \varepsilon_i \kappa^\perp_j}{\varepsilon_j \kappa^\perp_i + \varepsilon_i\kappa^\perp_j} \, ,\qquad  r_{s}^{i,j} = \frac{ \kappa^\perp_i -  \kappa^\perp_j}{\kappa^\perp_i + \kappa^\perp_j} \, ,
\end{equation}
respectively. In the non-retarded limit, the $z$-component of the  Casimir force density reads
\begin{equation}
	f(d)=-\frac{H}{6\pi d^3} \,,
\label{eq:result}
\end{equation}
with the Hamaker constant
\begin{equation}
    	H=\frac{3k_{\rm B} T}{2}\sum_{m=0}^\infty {}'\,\frac{\mathrm{Li}_3[r_p^+(i\xi_m) r_p^-(i\xi_m)]}{\varepsilon_{\rm L}(i\xi_m)}\,,
\end{equation}
 the polylogarithmic function
\begin{equation}
    \mathrm{Li}_3(y)=4\int\limits_{0}^{\infty}\mathrm d x \,x^2\frac{y e^{-2x}}{1-y e^{-2x}} \,,
\end{equation}
and the non-retarded reflection coefficients for $p$-polarised waves
\begin{equation}
    r_p^\pm = \frac{\varepsilon_\pm - \varepsilon_{\rm L}}{\varepsilon_\pm + \varepsilon_{\rm L}} \, .
\end{equation}
The estimation of the Hamaker constant via this type of experiments is very accurate, which decrease the errors of the reconstruction method as one can assume that the measured Hamaker constants are precisely measured. Current experiments reports on deviations below 5\% depending on the reduction of surface roughness, spatial resolution and error of the spring constant~\cite{PhysRevB.93.085434}.

\subsection{Mixing of fluids}
\begin{figure}[t]
\centering
 \includegraphics[width=0.6\columnwidth]{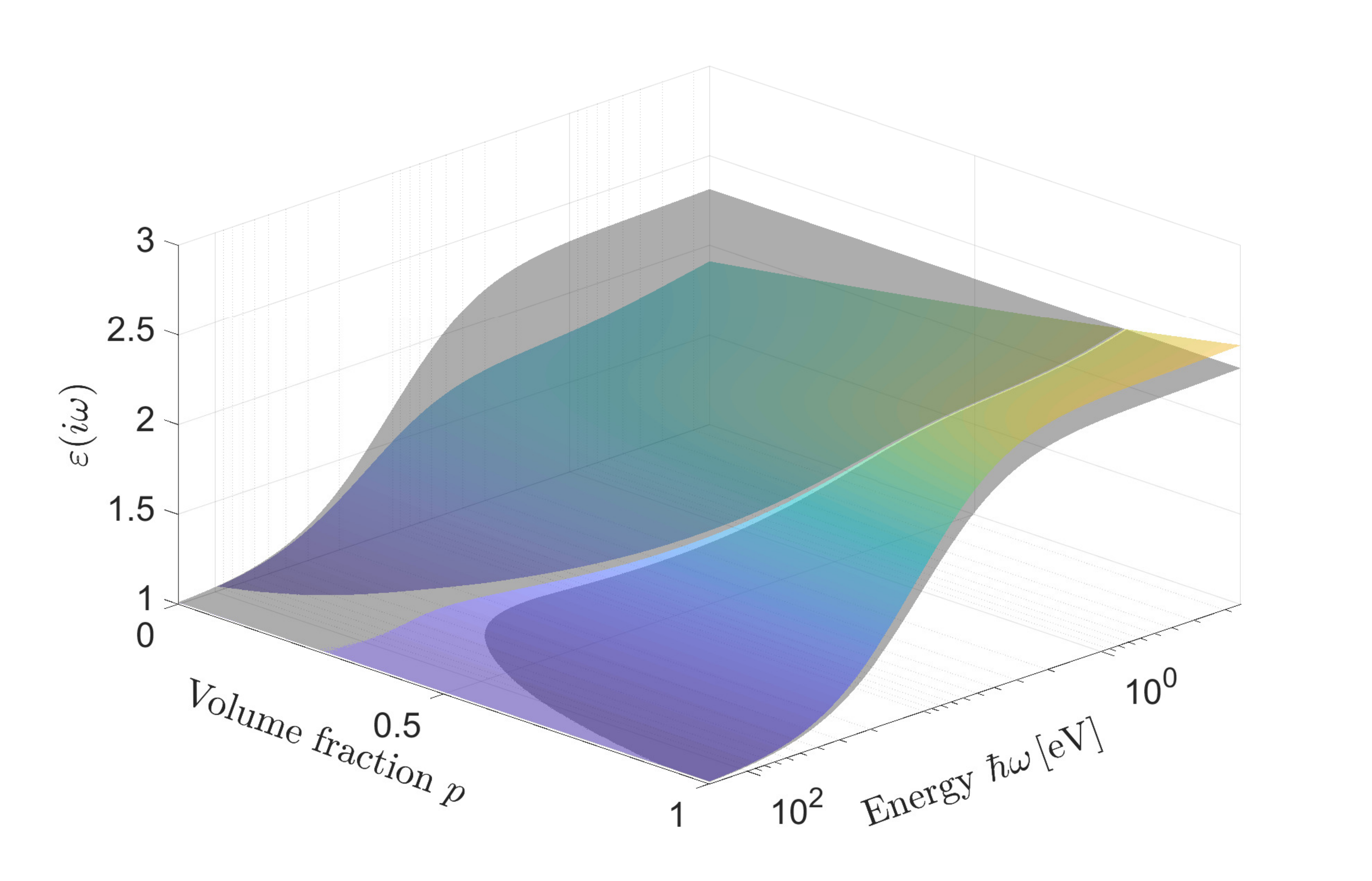}
 \caption{(Color online) Effective dielectric function $\varepsilon_{\rm L}(i\xi,p)$ of the mixed components methanol and bromobenzene depending on the volume fraction $p$. In order to illustrate the crossings in the dielectric function the grey surface denotes the permittivity of polystyrene $\varepsilon_-(i\xi)$.}\label{fig:polar2}
\end{figure}
In order to introduce tunable parameter, we consider a two-component fluid between both plates. For the dielectric functions of mixtures between fluid 1 ($\varepsilon_1$) in fluid 2 ($\varepsilon_2$)  we use a Lorentz--Lorenz like model~\cite{Aspnes}
\begin{equation}
\varepsilon_{\rm L}={\frac {1+2 \tilde\alpha}{1-\tilde\alpha}}\,,\quad \tilde\alpha= p {\frac {\varepsilon_{1}-1} {\varepsilon_{1}+2}}+(1-p) {\frac {\varepsilon_{2}-1} {\varepsilon_{2}+2}}\,,
\end{equation}
where $p$ is the volume fraction of fluid 1 in fluid 2. In order to illustrate the results, we choose the liquids bromobenzene and methanol in front of a polystyrene surface~\cite{PhysRevA.81.062502}. 
Figure~\ref{fig:polar2} illustrates the resulting dielectric function. One observes a change of the crossings of the dielectric functions depending on the volume fraction by following the intersection of the coloured surface (liquid) and the grey slightly transparent surface (polystyrene).

\subsection{Repeated measurements with changed concentration}
The result of repeated measurements with different concentrations is a series of Hamaker constants
\begin{eqnarray}
    	\lefteqn{H(p)=\frac{3k_{\rm B} T}{2}}\nonumber\\
    	&&\qquad\quad\times\sum_{m=0}^\infty {}'\frac{1}{\varepsilon_{\rm L}(p)}\mathrm{Li}_3\left[ \frac{\varepsilon_+ - \varepsilon_{\rm L}(p)}{\varepsilon_+ + \varepsilon_{\rm L}(p)} \frac{\varepsilon_- - \varepsilon_{\rm L}(p)}{\varepsilon_- + \varepsilon_{\rm L}(p)}\right]\,,\label{eq:Hp}
\end{eqnarray}
with $\varepsilon_+$ being the target dielectric function. In general, this function maps uniquely the target dielectric function onto the set of Hamaker constants, because reflection coefficients are typically bounded by $\left|r\right|\le 1$ which allows for the Taylor series expansion of the polylogarithmic function. 

\begin{figure}
    \centering
    \includegraphics[width=0.5\columnwidth]{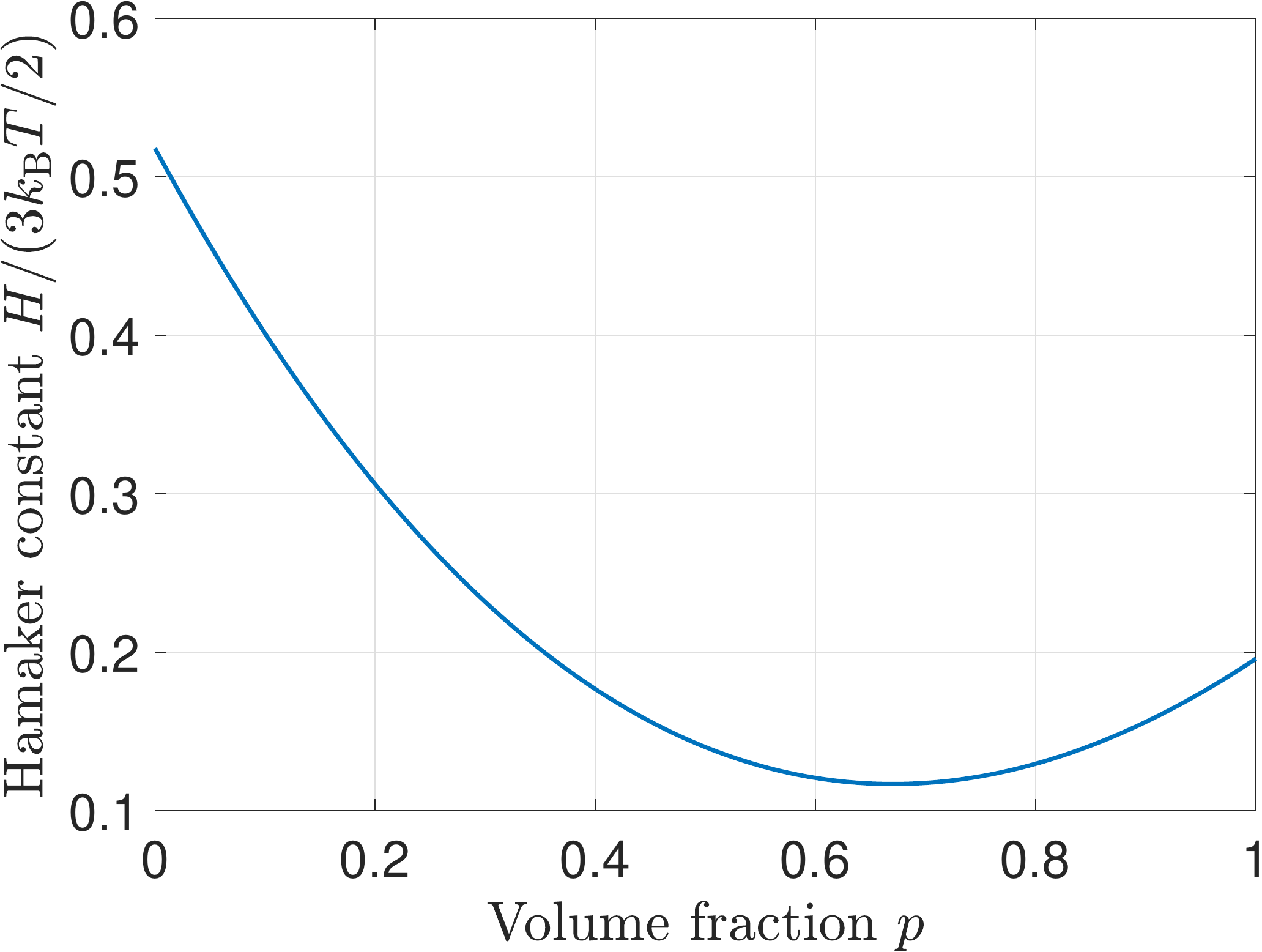}
    \caption{Example of Hamaker constants between two dielectric plates embedded in a two component liquid with the volume fraction $p$ giving the ration of bromobenzene in methanol.}
    \label{fig:hamaker}
\end{figure}
In order to illustrate the reconstruction method, we calculate the Hamaker constants for different volume fraction with a plate, where the target dielectric function to be reconstructed is given by a two-oscillator model
\begin{equation}
    \varepsilon_+ (i\omega) = 1+ \sum_i \frac{c_i}{1+(\omega/\omega_i)^2}  \, , \label{eq:EX}
\end{equation}
with $c_1 =2$, $c_2=1.5$, $\omega_1=2 \,\rm{eV}$ and $\omega_2 =6\,\rm{eV}$, which are typically values for a dielectric body.

\subsection{Reconstruction method}
The inverse problem is now defined as calculating the dielectric function of the plate $\varepsilon_+(\omega)$ given the knowledge about the Hamaker constants $H(p)$, the substrate dielectric function $\varepsilon_-(\omega)$ and the liquid dielectric function $\varepsilon_{\rm L}(p,\omega)$ via Eq.~(\ref{eq:Hp}). The dielectric function is a linear response function and has to satisfy specific properties, which will be exploited via the reconstruction method. First of all, the dielectric function is to satisfy causality, which leads to the connection of its real and imaginary parts via the Kramers--Kronig relation
\begin{eqnarray}
\mathrm{Re}\,\varepsilon(\omega) &=& 1+ \frac{2}{\pi} \mathcal{P}\int\limits_0^\infty \frac{\omega' \mathrm{Im} \,\varepsilon(\omega')}{{\omega'}^2-\omega^2} \mathrm d \omega'\,,\label{eq:KKR_Re} \\
\mathrm{Im}\,\varepsilon(\omega) &=&- \frac{2}{\pi} \mathcal{P}\int\limits_0^\infty \frac{\omega' \mathrm{Re} \,\varepsilon(\omega')}{{\omega'}^2-\omega^2} \mathrm d \omega'\,,
\end{eqnarray}
with the Cauchy principal value. By substituting $\omega=i\xi$ into (\ref{eq:KKR_Re}) the relation to imaginary arguments can be found
\begin{equation}
    \varepsilon(i\xi) = 1+ \frac{2}{\pi} \int\limits_0^\infty \frac{\omega' \mathrm{Im} \,\varepsilon(\omega')}{{\omega'}^2+\xi^2} \mathrm d \omega'\,. \label{eq:KKR_ixi}
\end{equation}
Further, the function is has to go to $1$ for large frequency arguments, $\lim_{\omega\mapsto \infty} \varepsilon(\omega) = 1$. Another important property is the analyticity in the upper complex half-plane. This fact together with the causality leads to the monotonic behaviour on the imaginary frequency axis. Due to the fact that the dielectric function is a linear response function sum rules exists that also have to be satisfied~\cite{AMBROSCHDRAXL20061} which are less important for our considerations as we neglect the damping of the response.

\begin{figure}
    \centering
    \includegraphics[width=0.6\columnwidth]{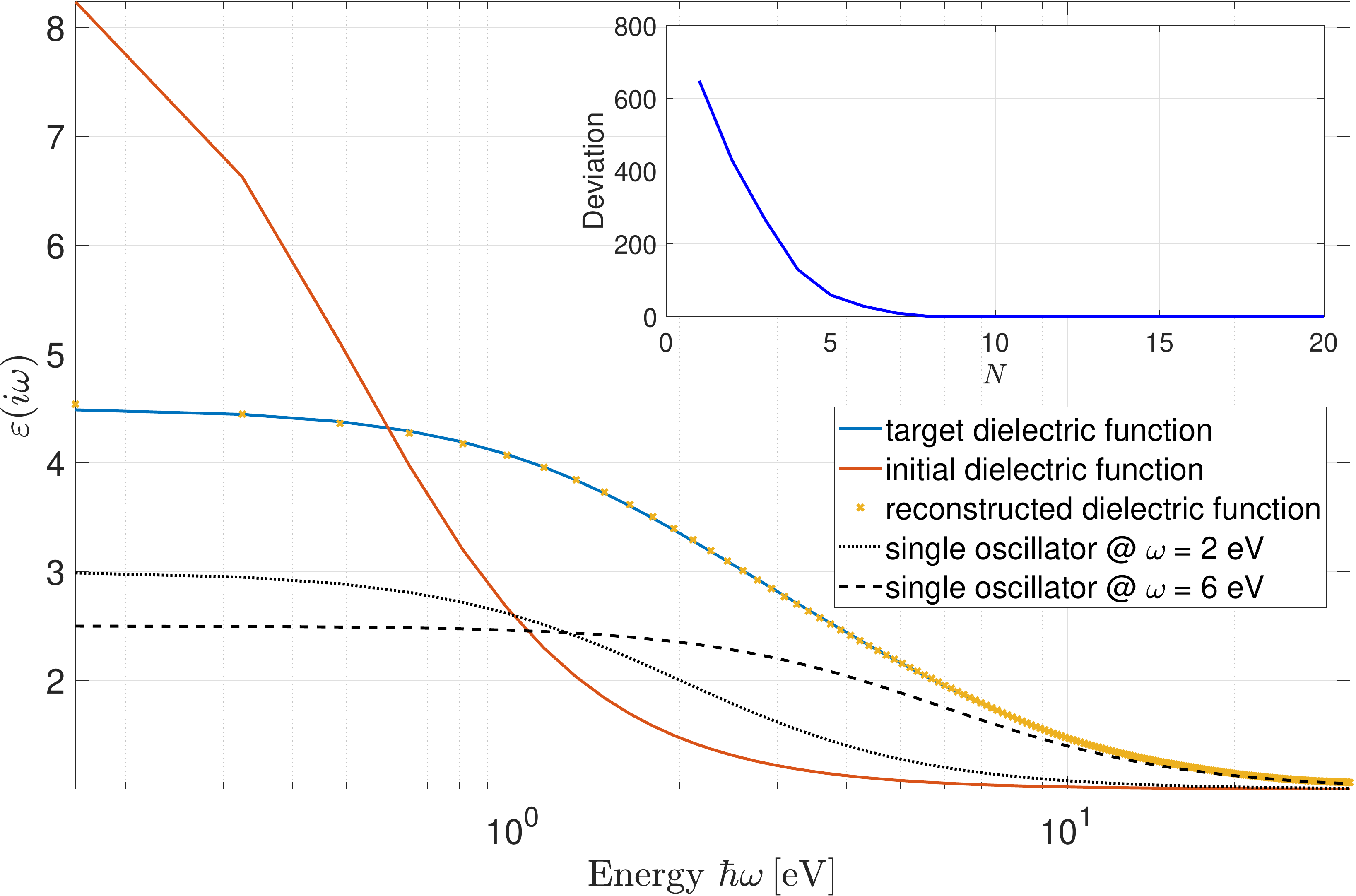}
    \caption{Agreement of the reconstruction method; the target dielectric function (blue solid line) is well-reproduced by the reconstructed dielectric function (yellow crosses). The initial dielectric function (red solid line) illustrates the convergences of the method. In the inlay plot the convergence of the method is depicted via the least square of the target function and the reconstructed function in each iteration step.}
    \label{fig:results}
\end{figure}

The Hamaker constant $H(p)$, which is depicted in Fig.~\ref{fig:hamaker} for the material combination with a polystyrene substrate and a bromobenzene and methanol liquid (both depicted in Fig.~\ref{fig:polar2}) and the sphere with dielectric function~(\ref{eq:EX}), depends non-linearly on the target function $\varepsilon_+$. Thus, we apply Newton--Raphson method to Eq.~(\ref{eq:Hp}), that transforms the the non-linearity into an iterative algorithm of systems of equations, which needs to be solved in the following way:
\begin{enumerate}
    \item[(i)] initialization, where we choose
\begin{equation}
    {{x}}_{0,i} = \varepsilon_+^{(0)}(i\omega_i) = 1+ \frac{A}{1+(\omega_i/B)^2} \, , 
\end{equation}
along a frequency grid of the first 200 Matzubara frequencies, with the parameters $A=8$ and $B=0.5\,\rm{eV}$. 
\end{enumerate} 
This initial function was chosen in order to ensure the monotonic decreasing of the dielectric function. The stability of the method strongly depends on the value $B$. For large values, the function overestimates the high-frequency regime too much leading to the convergence to a false (local) minimum. For small values, this region is strongly depressed, stabilising the method. In contrast, the chosen amplitude does not effect the stability of the method. Further, the static dielectric constant is overestimated by $\varepsilon_+^{(0)}(0) =9$. The target function to be determined only has $\varepsilon_+(0)=4.5$. Simulations showed that an overestimation of the static value is easier to handle for the algorithm. 
In order to run the method on an ordinary computer, we restrict ourselves to considering only the first 200 Matsubara frequencies (at room temperature), which covers the spectral range reaching extreme ultraviolet that covers most of the physical effects. This algorithm is followed by 
\begin{enumerate}
    \item[(ii)] the calculation of the matrix $D_{ij} = \frac{\partial H(p_i)}{\partial \varepsilon_{+}(i\omega_j)}$ to solve the system of equations
    \begin{equation}
        {\bf{D}} \Delta{\bf{x}}^{(i+1)} = H(p) - H(p,\Delta{\bf{x}}^{(i)}) \, ,
    \end{equation}
    with the measured Hamaker constants $H(p)$ and the calculated Hamaker constants based on the previously iterated data according to Eq.~(\ref{eq:Hp}).
\end{enumerate}
The matrix ${\bf{D}}$ typically has singularities. To overcome this issue, we applied the Moore-Penrose Pseudoinverse. It is known that Newton's method converges locally, which means that it strongly depends on the initial data. This effect is directly observable by  oscillations in the target function that violates the causality properties of the dielectric function mentioned above.  In order to solve this issue, we regularly force the newly iterated function onto a causal monotonic profile by
\begin{enumerate}
    \item[(iii)] fitting the result to a strictly monotonically decreasing function
    \begin{equation}
        {{x}}_j^{(i+1)}={{x}}_j^{(i)}+\Delta{{x}}_j^{(i)} \approx 1+\frac{\lambda_1}{1+(\omega_j/\lambda_2)^2} \,, \label{eq:fit}
    \end{equation}
    with the parameters $\lambda_{1,2}$.
\end{enumerate}
We find that a fitting to this single oscillator model, which satisfies the monotonic behaviour, works optimal and reproduces the more complicated two-oscillator target after the next iteration. This favourable behaviour is caused by the large differences between the oscillator strengths in the example~(\ref{eq:EX}). On smaller scales with two or more narrow peaks, we introduce a termination condition for the filter (iii) according to the iterated step by 
\begin{enumerate}
    \item[(iv)]Calculate the maximum iterative step size 
    \begin{equation}
        \max_j \left|\Delta\varepsilon_+(i\omega_j)\right|<C \, , \label{eq:condition}
    \end{equation}
    which we chose in our case to be $0.1$. By step sizes below this threshold, the fitting step (iii) will be left out.
\end{enumerate}
 We also tested the reconstruction method for models with more oscillators and the data were reproduced with less than  100 iterations. The result of this reconstruction method with the target function~(\ref{eq:EX}) is depicted in Fig.~\ref{fig:results}. It can be observed that the target function is reproduced 6 iterations. Especially, the regions around the two oscillators (at $2 \,\rm{eV}$ and $6 \,\rm{eV}$) are well reproduced. A small deviation of the static value is observed. The presented inversion method has some limitations according to the possible reconstruction of completely unknown dielectric functions. In the presented way the target dielectric function has to be quite smooth in order to be approximated by a single oscillator response function with respect to the condition~(\ref{eq:condition}). In order to take errors due to this condition into account one needs the redo the calculation with different values. Due to a generalisation of this method to arbitrary response functions, one needs to record the oscillating profiles in between where one observes fix points that already match the target function. Thus, an improvement of the method can be performed with respect to two possibilities: (a) by optimising the fitting in step (iii); or (b) by reducing the system of equations, step (ii), to reproduce these values directly. 

After the reconstruction of the target dielectric function on the imaginary frequency axis $\varepsilon_+(i\xi)$, the result needs to be transformed to the real frequency axis. This can be done via the integral equation~(\ref{eq:KKR_Re}) which is an inhomogeneous Fredholm equation of the first kind. This procedure results in the imaginary part of the dielectric function and after the additional application of the Kramers--Kronig relation~(\ref{eq:KKR_Re}) its real part can be obtained.

\section{Conclusions}
We have introduced a reconstruction method to estimate the dielectric function of a solid plate via the established Casimir experiments.  In order to estimate the dielectric function along 200 Matsubara frequencies, 200 Hamaker constants at different concentrations are required. 
For larger distances, the frequency and spatial dependences of the interaction do not separate~\cite{Fiedler_2017} which can be exploited to reduce the number of experimental repetitions. 
A more efficient reduction of experiments towards an application of this method can be found via the training of a neuronal network, which will be object of investigation in further studies. Beside the introduced application of the reconstruction of the dielectric function of the attached nanoparticle, the method can directly be used to measure the substrate and after some modification to measure the dielectric function of the liquid. The decomposition of the liquid or gas components will also be part of future investigations. Further, this method can be adapted to low spectral ranges, terahertz radiation and below, where current methods fail, by combining the proposed method with experimentally obtained data from higher energy ranges. By cooling the system down the to several Kelvin, the natural discretisation yields roughly 100 points covering the terahertz regime. Further cooling will lead to the transition to the zero-Kelvin Casimir force, where the discrete sum has to be exchanged by the continuous integral which results in a free parameterisation for the investigated regime depending on the knowledge of further optical data. Experimentally challenging is the control of two liquids at the required low temperatures, which can be done by using superfluids or ultracold quantum gases. 

\section*{Acknowledgments}
We acknowledge support from the Research Council of Norway (Project  250346). We gratefully acknowledge support by the German Research Council (grant BU 1803/6-1, S.Y.B. and J.F., BU 1803/3-1, S.Y.B.),  the Freiburg Institute for Advanced Studies (S.Y.B.).

\bibliographystyle{unsrt}  
\bibliography{bibi}  


\end{document}